# Effect of gamma radiation on electrical properties of diffusive memristor devices

D.P. Pattnaik[1*], C. Andrews[2], M. D. Cropper[1], A. Balanov[1], S. Savel'ev[1], and P. Borisov[1]

*1. Physics Department, Loughborough University, Loughborough, LE11 3TU, United Kingdom*

*2. University of Manchester, Dalton Cumbrian Facility, Westlakes Science Park, Moor Row, CA24 3HA, United Kingdom*

E-mail: d.pattnaik@lboro.ac.uk

Keywords: memristors, diffusion, nanoparticles, gamma radiation, neuromorphic

**Abstract**

Diffusive memristors continue to receive tremendous interest due to their ability to emulate biological neurons and thus aid the development of bio-inspired computation technology. A major issue with the diffusive memristor is the inability to reliably control the formation of the conduction filaments which affects both the device functionality and reproducibility of regimes after each application of voltage. Here we investigate the effect of gamma radiation on the electrical properties of the diffusive memristors based on metallic nanoparticles in dielectric matrix. Our experiments show that after exposing to radiation, the memristors demonstrate much sharper (and less noisy) hysteresis in the current-voltage characteristics while preserving the same low- and high-resistive states as in the pristine samples. Additionally, the radiation lowers both threshold and hold voltages that correspond to onset of low- and high- resistive states, respectively. The proposed mechanism involves radiation-induced defects in the silica matrix which help to establish dominant pathways for nanoparticles to form conduction filaments. Our findings suggest an efficient way to enhance working characteristics of diffusive memristors and to improve their reproducibility.

**1. Introduction**

Abilities of memristors to change their resistance depending on current or voltage history offer tremendous potential for the development of the next generation of non-volatile memory such as Resistive Random-Access Memory (ReRAM) and brain-inspired neuromorphic hardware. [1, 2, 3, 4] The fundamental property of these devices is an I-V hysteresis loop, where the memristor changes its resistance at certain voltage thresholds. For a diffusive memristor, [5, 6] the composite material is made of a dielectric matrix, typically $SiO_x$, $MgO_x$, or $HfO_x$ with embedded metallic nanoparticles (NPs) of Au, Ag or Cu [7] which are either formed as a result of co-deposition in a single thin film or by growing a thin metallic layer next to the dielectric one. When an external voltage is applied between the electrodes, the NPs diffuse in the dielectric matrix and form a conduction filament (CF) between the electrodes (SET). When the external voltage is removed, the CF collapses (RESET) due to minimization of interfacial surface energy. The formation and rupture of the CF manifests as a change between high resistance state (HRS) and low resistance state (LRS), with corresponding voltages labelled as threshold voltage $V_{th}$ and hold voltage $V_h$ respectively. [7, 8, 9, 10] Mechanisms behind filament formation and rupture are of critical importance for design of neuromorphic hardware based on diffusive memristors. [2, 11, 12, 13, 14]

Diffusive memristors made of $SiO_2$ layers are structurally similar to a typical CMOS (complementary metal-oxide-semiconductor) device. Previous works on exposure of CMOS and oxide-based ReRAM devices to different sources of ionizing radiation demonstrate that its effects on the device properties are rather complex, however, in general, they could be reduced to generation of electron-hole pairs in the oxide layer. A typical hole yield for [60]C gamma photons in $SiO_2$ in zero-field was found to be 0.3 if the subsequent recombination has been considered. [15, 16] In $SiO_2$ devices ionizing radiation creates broken bonds between silicon and oxygen. [17, 18, 19, 20] A broken Si-O bond can yield a trivalent silicon with a dangling bond and a non-bridging oxygen which can further release an electron-hole pair or a hole only while the non-bridging oxygen transforms into to a hole trap state:

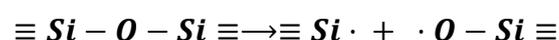

$$\equiv Si - O - Si \equiv \longrightarrow \equiv Si \cdot + \cdot O - Si \equiv$$

$$\equiv Si\cdot + \cdot O - Si \equiv \rightarrow h^+ + e^- + \cdot O - Si \equiv$$

$$\equiv Si\cdot + \cdot O - Si \equiv \rightarrow h^+ + O^- - Si \equiv$$

In SiO$_2$ the electrons usually drift away from the original point of generation within picoseconds due to their high mobility, [21] while the less mobile and heavier holes remain inside the oxide and cause the effective reduction of the positive voltage bias when it is applied. However, over time, which could be from seconds to years, depending on the device geometry, temperature and field environment, the holes will also drift away causing a short-term recovery towards the original state. When these holes reach long-lived trapped states, typically near the oxide interface, they build up trapped charge that can cause a more long-lasting change to the device performance ranging from few hours to years. [16] Further, exposure to ionizing radiation also leads to formation of variety point defects beyond trivalent silicon and non-bridging oxygen such as oxygen vacancies, silicon vacancies, interstitial oxygen, and interstitial silicon. These act as dopants and trap states, usually forming positive space charge in the oxide layer. [22, 23, 24, 25, 26] However, for the purpose of resistive switching those defects can have significant impact on device properties.

In case of non-volatile resistive switching in ReRAM thin layers of Cu- doped SiO$_2$ [27] or HfO$_2$ [28] with Cu and Pt or Cu and W electrodes, it was reported that formation and rupture of CFs is only weakly affected by various doses of $^{60}$Co γ radiation (total dose of 3.6 kGy [28] or 71 kGy [27]), that is the resistances in LRS, reset voltages, and the switching endurance exhibited no significant reduction, and the SET-RESET process remained reversible. Only the HRS resistance and the set voltage values showed some relatively small decrease and increase, respectively, after the irradiation. Yuan et al, [29] reported for devices with Ag doped AlO$_x$ layers demonstrating non-volatile resistive switching that the LRS resistance, set and reset voltage values were almost stable (only slight decrease in set voltage values) upon γ-irradiation from a $^{60}$Co source at the total dose values of 5kGy and 10 kGy. However, the HRS resistance and forming voltage values needed to perform the initial electroforming were found to decrease and increase, respectively. This was interpreted, firstly, as a result of increased migration and spatial dispersion of Ag ions into the dielectric matrix after the irradiation, and secondly, as a signature of radiation induced holes which became trapped in the AlOx layer near the bottom electrode (due to the work function difference) opposite the top electrode with larger concentration of Ag NPs. This caused an increase in the forming voltage values as a higher electric field was needed to initiate field-driven Ag diffusion over increased barriers and trapped charges in order to form the initial switching filament during the electroforming process. Further it was suggested that the trapped holes facilitated formation of Ag CFs after the electroforming, hence some decrease in the set voltage values was observed. At the same time, this was reported to be in contradiction to findings in for a Cu-doped $HfO_2$ system where the opposite was found. [27] At the same time, both Refs [29] and [28] found a similar decrease in HRS resistance, which was explained in Ref. [29] by the increased leakage due to tunneling through new radiation induced defects.

In this letter, we studied the impact of γ-radiation from a $^{60}$Co source with a total dose of 50 kGy, on diffusive memristor devices made of a $SiO_2$ layer doped with Ag. Our measurements revealed that the exposure to radiation led to more sharp hysteresis in I-V curves of the devices and to decrease of the corresponding threshold voltages. We also found improvement in stability of resistive switching. Further, the artificial spiking neurons made of the irradiated devices demonstrated much higher spiking frequencies in comparison to pristine samples, which is promising for acceleration of neuromorphic computations.

## 2. Results and Discussion

### 2.1. I-V Characteristics

I-V curves of four devices from the same batch were measured and averaged after five voltage sweeps as shown **Figure 1 (a)-(d)** before and after the irradiation. A noticeable change in memristor switching behavior can be observed in all the four devices after the irradiation. For the pristine samples the switching between HRS and LRS was much more gradual and included numerous irregular current jumps, likely due to randomness in CF formation and destruction with increasing or decreasing the applied voltage. At the same time, the device current demonstrated significant noise which we explain by instabilities in the CF.

In contrast to this, after being exposed to radiation the same devices started to demonstrate more distinct and stable threshold switching behavior. For the increasing voltage sweep, the resistance switching from HRS to LRS became rather abrupt and happened at lower threshold voltage values, which suggests that CF formation must have required less time and energy after the irradiation. The reset sweep showed similar abrupt switching behavior from LRS to HRS. The more stable and less fluctuating I-V curves allow also to assume more reproducible and less random CF formations. These effects could be explained by appearance of the radiation-induced defects in the dielectric matrix that promote nucleation of the CF and direct it further formation.

2.2. XPS Analysis

The chemical state and composition of the device switching layer before and after the irradiation were studied by XPS measurements. Note that the XPS is only sensitive to the top few nm of the film, however, we expect no significant difference in the film chemical composition across the film thickness. The XPS spectrum for Ag taken on the sample before irradiation (**Figure 2(a)**) shows $3d_{3/2}$ and $3d_{5/2}$ peaks corresponding only to metallic Ag, however, after irradiation each peak became split into two corresponding to metallic Ag and silver oxide phases (**Figure 2(b)**), that is, partial silver oxidation must have taken place after the irradiation. [27, 31, 32, 33]. This is further corroborated by the change in the position, shape

and width of the Auger MNN Ag peaks measured before and after irradiation (**Figure 2 (c)**). [32, 34]

A typical XPS spectrum for Si 2p taken before irradiation (**Figure 3(a)**, black line) demonstrates a single peak corresponding to silicon oxide. After irradiation (**Figure 3(a)**, red line) its height significantly decreases accompanied by a slight shift in the peak position, while an additional, new peak corresponding to pure Si appears at 96.78 eV [35] evidencing some reduction of oxidized Si. Similarly, the XPS spectrum for O 1s (**Figure 3 (b)**, black line) shows a single peak before the irradiation, corresponding to oxygen in silicon oxide, which decreases in the intensity after the irradiation (**Figure 3 (b)**, red line) and becomes accompanied by a new additional peak at 530.38 eV [35, 36] which is assigned to a metal-oxide phase (in this case likely silver-oxide). This combined evidence from the Si 2p peak shift along with the appearance of the pure Si and metal-oxide O1s peaks after the irradiation supports a hypothesis that some Si in $SiO_2$ was reduced and that some oxygen defects in the form of oxygen vacancies are introduced in the film due to the irradiation. [37, 38]

Further, comparison of XPS spectra before and after the irradiation implies that some amount of interstitial oxygen must have been generated thus partially oxidizing the silver clusters and leaving oxygen vacancies in silica dielectric matrix. Here, it is natural to assume that some silver ions could have reacted with the non-bridging oxygen at the interface to $SiO_2$. In parallel to formation of interstitial oxygen, generation of interstitial silicon must have happened due to many broken Si-O bonds which should release some amount of silicon out of the crystal lattice.

2.3. Artificial Neuron Spiking

For further insight into the effects of γ-radiation on our devices we investigated the spiking behavior of the artificial neurons whose electrical circuit is shown in **Figure 4(a)**. In our study, the artificial neuron included a diffusive memristor connected in series with an external resistor

$R_L$= 55 kΩ and in parallel to an external capacitor $C_P$=1 nF. This circuit was powered by a constant voltage $V_{ext}$=1V. The applied $V_{ext}$ charged $C_P$ until the voltage drop across the memristor reached its threshold value causing switching to LRS. When in LRS, the capacitor discharged producing a current spike whilst the device voltage has decreased until it became below the hold voltage and the device switched back to HRS. This process of repeated charging and discharging continued itself and resulted in a series of electric current spikes mimicking spiking of biological neuron. [7]

Comparison of voltage spiking in an artificial neuron with the same memristor measured before and after the irradiation (**Figure 4 (b)**, top and bottom panel, respectively) shows a dramatic increase in the spiking frequency after the irradiation. As the spiking frequency is determined by the rate of formation and rupture of the CFs, this increase means that the radiation-induced defects promote faster and more reliable formation and disruption of CFs.

**3. Mechanism**

3.1. Hypothesis

As discussed previously, the transition from HRS to LRS in diffusive memristors made of $SiO_2$ doped with Ag is realized by Ag NPs clustering in a CF between the two electrodes driven by the external electric field as illustrated in **Figure 5 (a).**

After γ- irradiation, the oxide layer comprising Ag NPs is likely to contain high concentration of trapped holes, oxygen vacancies, interstitial oxygen, and interstitial silicon. The XPS data, described above suggests that after irradiation, at least some fraction of the interstitial oxygen does oxidize interfacial silver ions, whilst the interstitial silicon atoms together with oxygen vacancies are left distributed in the oxide layer. These silicon ions could consequently diffuse to form silicon nano inclusions (NI) similar to the ones reported in [39, 40]. Some redistribution of interstitial silicon and oxygen vacancies could have also happened due to work function difference (caused by the difference in electrode roughness [14] or film thickness [39, 41, 42] ) between the electrodes, however, its effect for the device performance is not that significant. The subsequent field application should further produce oxygen vacancies to concentrate at the electrode opposite the positive one.

We hypothesize that the metastable structure consisting of silicon rich silica which readily segregates to oxygen vacancies along with the $Si^+$ contribute to the modification of the electrical properties of our diffusive memristor by creating a prevalent guiding conduction path for Ag NPs. [43] Consequently, Si NI clusters emerged after irradiation enhance and accelerate formation of CF switching the device to LRS, see **Figure 5(b)**. Namely, Ag NPs do not need any more to form a continuous CF connecting the top and bottom electrodes but instead to amass a series of smaller CFs between the established Si NI clusters within the oxide layers. Thus, the energy barrier to form a LRS is lowered resulting in reduction of $V_{th}$ and the dynamics of CF formation happens on shorter timescales which explains more abrupt threshold switching in the IV curves and higher spiking frequency in artificial neurons made after the irradiation. The distributed Si NIs, perhaps with support by oxygen vacancies, form a backbone on which a conduction pathway is formed as a series of CFs stretching from one electrode to another. As the result, the formation (and disruption) of conducting channels with variation of $V_{ext}$ become more stable and reproducible. Moreover, increased doping of silicon oxide with holes and electrons could promote tunneling between Ag NPs which do not form part of a CF, and between broken filaments where a continuous conduction is not possible for some reason. However, more rigor verification of the role of Si NIs as well of the effects of trapped charges and oxygen vacancies in the oxide requires further theoretical and experimental studies. [39, 44, 45]

3.2. Theoretical Model and Simulations

In line with the above hypothesis, the influence of radiation on the filament formation effectively constrains particle drift and diffusion. Before the irradiation, almost all volume between memristive terminals is available for Ag clusters, and after the irradiation pinning centers for Ag clusters are created, thus forming preferable channels/valleys for Ag NPs to move between terminals. Regardless of the character of the interaction between the pinning centers and the Ag NPs, that is, whether it is attractive or repulsive, the room for the particle diffusion is effectively shrinks towards certain preferable paths. Such a constraint could be considered as a transition from 2D diffusion [46] to the so-called "single-file diffusion" [47]. The effective shrinking of the transfer space can be modelled by varying the transverse trapping potential within the commonly accepted diffusive memristor model (see supplementary information in Ref. [48]).

Within the model, two component random force ($\xi_{i,x}$, $\xi_{i,x}$) with zero mean ($\langle\xi_{i,x}\rangle$, $\langle\xi_{i,x}\rangle$)=(0, 0) and delta-correlations in time $t$, $\langle\xi_{i,x}(0)\xi_{i,x}(t)\rangle = \delta(t)$, $\langle\xi_{i,y}(0)\xi_{i,y}(t)\rangle = \delta(t)$, $\langle\xi_{i,x}(0)\xi_{i,y}(t)\rangle = 0$, together with the drift force $\frac{qV}{L}$ ($q$ is charge of Ag cluster, $V$ electric voltage, $L$ is the gap between the electrodes or between arms of forming filament) control diffusion of $i$-th Ag NP with the coordinates $(x_i, y_i)$ in a two-dimensional potential $U(x_i, y_i) = U_x(x_i) + \alpha y_i^2$. The potential along $x$-axis $U_x(x_i)$ (see inset in **Fig 6 f**) has a large minimum nearby one of the memristor terminals reflecting attraction to Ag-rich areas to minimize Ag-SiO interface energy, while the parabolic potential $\alpha y_i^2$ implies constrain for transverse particle diffusion, that is, the one that is controlled by the presence of radiation-induced pinning centers, i.e. the larger $\alpha$ the more single-filed the diffusion becomes.

$$\eta\frac{dx_i}{dt} = -\frac{\partial U(x_i,y_i)}{\partial x_i} - \sum_{j\neq i}\frac{\partial W(x_i-x_j,y_i-y_j)}{\partial x_i} + q\frac{V}{L} + \sqrt{2\eta k_B T}\xi_{i,x},$$

$$\eta\frac{dy_i}{dt} = -\frac{\partial U(x_i,y_i)}{\partial y_i} - \sum_{j\neq i}\frac{\partial W(x_i-x_j,y_i-y_j)}{\partial y_i} + q\frac{V}{L} + \sqrt{2\eta k_B T}\xi_{i,y},$$

$$\frac{dT}{dt} = -\frac{V^2}{CR_M} - \kappa(T - T_0),$$

$$\tau \frac{dV}{dt} = V_{ext} - \left(1 + \frac{R_L}{R_M}\right)V.$$

Here, $T$ is the cluster temperature, which can significantly be different from the memristor matrix temperature; $T_0$ is the bath temperature (e.g., SiO-matrix or memristor terminals/substrate temperature), $\eta, k_B$ are Ag-cluster viscosity and the Boltzmann coefficient, while $\kappa, C$ are the heat transfer coefficient and cluster heat capacity, respectfully. Repulsive interaction of Ag-clusters defined by the potential $W = W_0 \exp(-\rho_{i,j}/r_{int})$ with $\rho_{int} = \sqrt{(x_i - x_j)^2 + (y_i - y_j)^2}$ prevents their agglomeration in the potential well and controls the transition from 2D to single file diffusion. [49, 50] The last equation in the set is the Kirchhoff's voltage law for artificial neuron circuit with voltage bias $V_{ext}$ applied to the artificial neuron, the load resistance $R_L$, parallel capacitance $C_P$, and RC time constant $\tau$, that is, the model describes spiking behaviour of our artificial neuron before and after the irradiation (see Figure 4 for the circuit scheme and experimental data). The resistance of the memristor is calculated by considering all paths between the electrodes through Ag-clusters and assuming tunnelling resistance between two clusters $i$ and $j$, $R_{i,j} = \exp\left(-\frac{\rho_{i,j}}{\lambda}\right)$ if several clusters are missed. The resistance of each path $p$ is estimated as a series connection of tunneling resistances of all clusters in that path, and then the total memristor resistance $R_M$ is evaluated as a parallel connection of all contributing paths, such that the total conductance is:

$$G_M = \frac{1}{R_M} = \sum_p \left(\sum_{i,j \in p} R_{i,j}\right)^{-1}$$

As was shown earlier, such ''compact models'' [51] allow to qualitatively describe all observed experimental features and provide an insight into underlying physical phenomena. In our simulation for simplicity, we consider a case of four particles.

The dynamics of each of the four Ag particles was simulated for two cases: when lateral diffusion is high (**Figure 6 a, c, and e**, $\alpha = 1$), representing the sample before the irradiation; and when lateral diffusion is suppressed (**Figure 6 b, d** and f, $\alpha = 100$), representing the sample after irradiation. Before the irradiation, the conductance demonstrates a very complex irregular dynamics (Figure 6a) driven by rather erratic behavior of Ag-clusters. Collective movement of different Ag particles along the x-axis (Figure 6c, different colors) is characterized by them overtaking each other frequently which imposes randomness into the total conductance, e.g., the Ag-cluster shown by black curve overtakes Ag clusters with other trajectories at $t \approx 0.5$. This randomness is also reflected by random walks of the particles in the *y*-direction as illustrated in Figure 6 e. For the sample after irradiation the fluctuations in the *y*-direction are significantly subdued (Figure 6 f), evincing the transition from a 2D to a quasi-1D dynamics. This also makes the movement of Ag-clusters in the *x*-direction (see Figure 6 d) more regular, as those start to move one after another with no overtaking taking place. Such ordered collective behavior of the Ag-clusters accelerates the conductance spiking, which becomes not only more regular, but also more frequent (cf. **Figure 6a and b**). Thus, these results are in a perfect qualitative agreement with experimental measurements of artificial neuron spiking presented in Figure 4b, which justifies a transition from 2D to quasi-1D drift-diffusion after irradiation as a plausible physical mechanism.

## 4. Conclusion

In conclusion, we experimentally and theoretically studied effects of γ- radiation from a $^{60}C$ source on electric properties of Ag -based diffusive memristors. The devices after irradiation demonstrated lower threshold voltages and more abrupt resistance switching between HRS and LRS. We also found out that artificial neurons with the memristors after irradiation demonstrated higher frequency and more regular spiking than those with pristine memristors.

These phenomena can be explained by formation of radiation induced Si- nano inclusions which direct and accelerate formation of CFs. Our simulations suggest that the physical mechanism of the observed dramatic change of spiking is the dynamical transition from 2D complex diffusion in the pristine samples to single-file diffusion in the samples being exposed to radiation. Our findings not only shed a light on how an exposure to a high energy radiation affects the charge transport in diffusive switches, but also offer an efficient way to improve the performance of diffusive memristors and the related artificial neurons based on these memristors via creation of artificial pinning centers. They also stimulate further research in understanding the roles of various radiation-induced defects and nano-inclusions on charge transport of memristive devices, which would promote development of more endurable and radiation immune technological concepts.

## 5. Methods

*Sample preparation*: A bottom electrode of 70 nm Pt was deposited by magnetron sputtering on a $SiO_2$/Si wafer, followed by co-sputtering of 60 nm of Ag and $SiO_2$ in a mixed atmosphere of Ar and $O_2$. A top electrode of 30 nm Pt was sputtered through a shadow mask with circular holes of 100 μm in diameter to obtain the following film structure: $SiO_2$\Pt\$SiO_x$:Ag\Pt. All the layers were deposited at room temperature of the substrate and at a growth pressure of 5.5 mTorr.

*Electrical characterization*: Device contacts were made using tungsten tips housed in a probe station by Everbeing. Current-voltage characterization was performed using a Keithley 4200 SCS parameter analyzer. For measurements of self-sustained current spikes, a voltage pulse (1V, 50s) was applied to the device using a Rigol waveform generator and the device voltage was recorded using a PicoScope digital oscilloscope whilst the memristor was connected in series to a load resistance $R_L$= 65 kΩ and in parallel to a capacitor $C_P$=1 nF.

*X-ray photoelectron spectroscopy*: XPS was performed using a Thermo K-Alpha system with an Al Kα mono-chromated (1486.6 eV) source with an overall energy resolution of 350 meV. The XPS peaks are charge corrected to adventitious carbon at 284.8eV.

*Radiation exposure*: The gamma radiation exposure of the devices was carried out at the Dalton Cumbrian Facility using an irradiator in a self-contained Foss Therapy Services Model 812 with a 9 L sample chamber. The samples received $^{60}$Co gamma (γ) radiation with two energies of 1.17 MeV and 1.33 MeV (average energy of 1.25 MeV) at a dose rate of 229.51 Gy/min for exactly 217.86 minutes to give a total accumulated dose of 50 kGy. This dose was chosen as previous works on other memristive systems had reported no significant change to electrical properties for lower radiation doses. [24, 29] The irradiator contained three source rods with up to three source capsules (GIK-7M-4) with an initial activity of 2500 Ci per rod such that the activity was evenly distributed along the length of each source rod.

*Theoretical simulation:* Simulation parameters are: $\frac{\lambda}{L} = 0.1, \frac{r_{int}}{L} = 0.04; \frac{W_0}{U(-1)-U(0)} = 0.6; \kappa\tau = 90; \alpha L^2/(U(-1) - U(0)) = 1$ and 100 for 2D and single-file diffusion, respectively; $\frac{qV}{U(-1)-U(0)} = 60; \frac{R_L}{min}R_M = 300; T_0/(U(-1) - U(0)) = 0.0012$; we use units where $2k_B = 1, \eta = 1$, C=1.

**Acknowledgements**

The authors would like to thank Sam Davis for the XPS characterization and acknowledge the use of the facilities within the Loughborough Materials Characterisation Centre. The authors acknowledge the support of The University of Manchester's Dalton Cumbrian Facility (DCF), a partner in the National Nuclear User Facility, the EPSRC UK National Ion Beam Centre and the Henry Royce Institute. We acknowledge Ruth Edge for the assistance during the Gamma irradiation. This work was supported by The Engineering and Physical Sciences Research Council (EPSRC), grant no. EP/S032843/1.

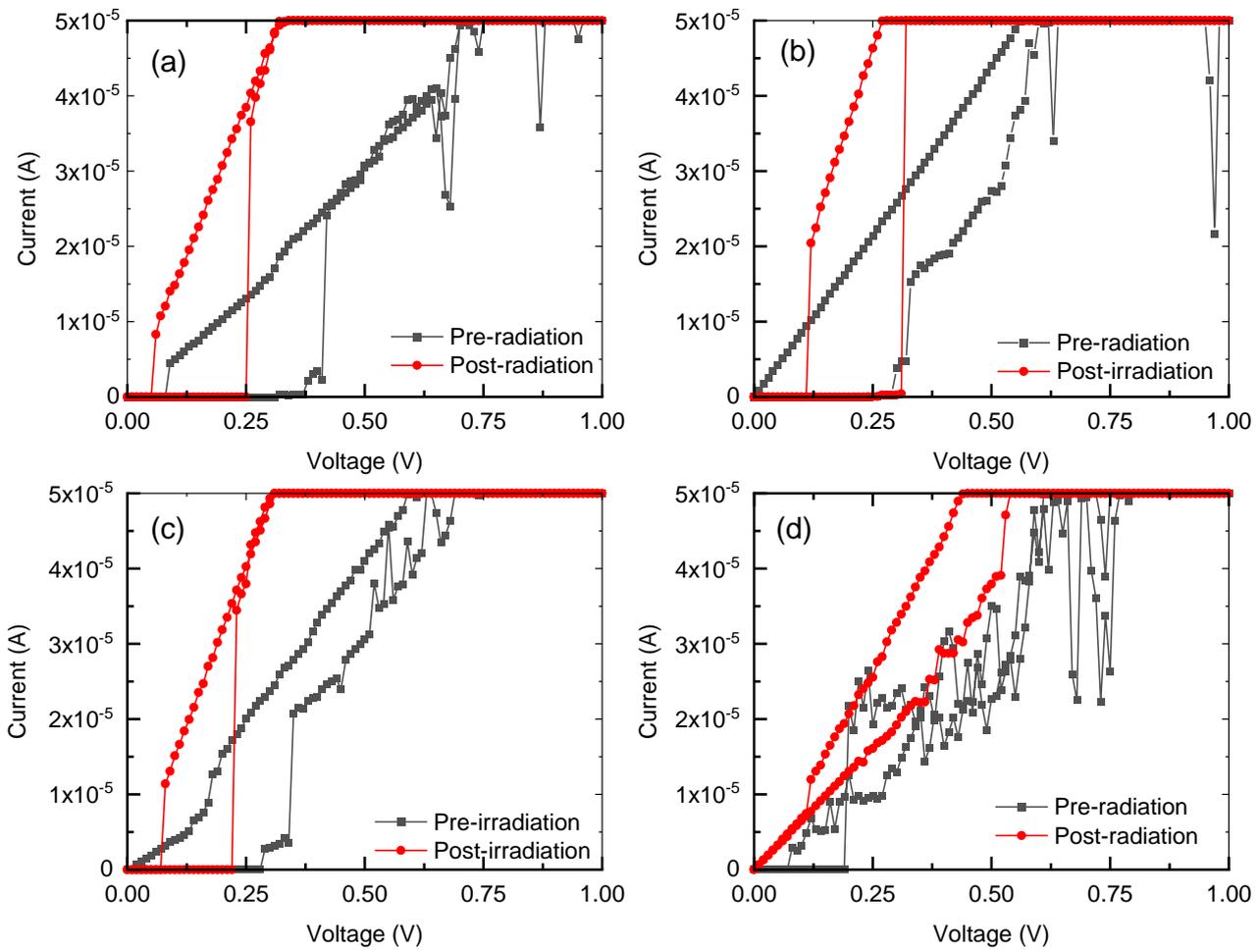

**Figure 1.** (a)-(d) Current-voltage characteristics for four separate devices from the same batch before (black) and after the irradiation (red).

**Figure 2.** (a, b) XPS spectra (black line) for Ag $3d_{3/2}$ and $3d_{5/2}$ peaks before (a) and after the irradiation (b). The corresponding fits represent peaks for metallic Ag (red line) and silver oxide (blue line) in $3d_{5/2}$. **(c)** Auger MNN peak for Ag before (red) and after irradiation (blue).

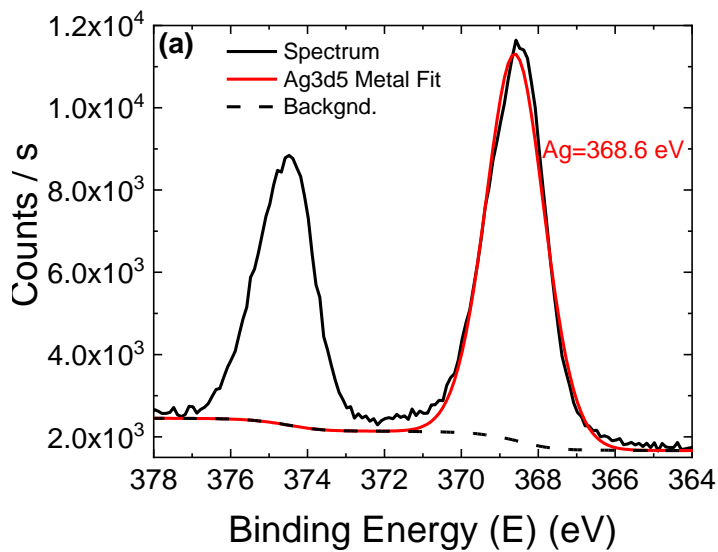

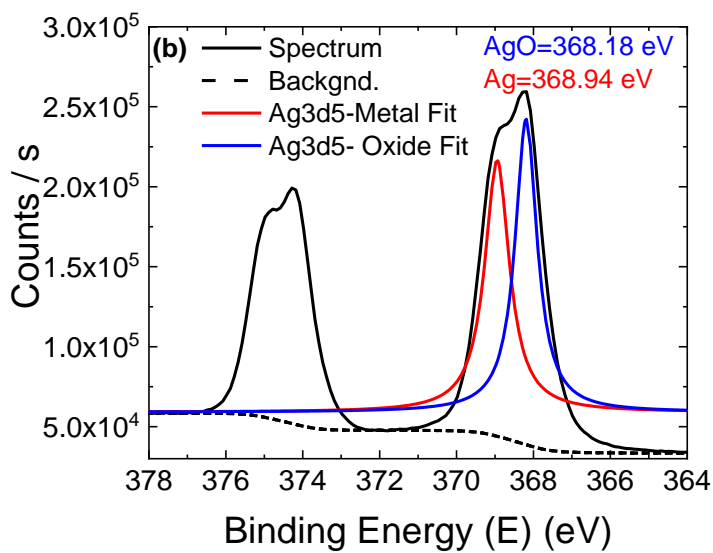

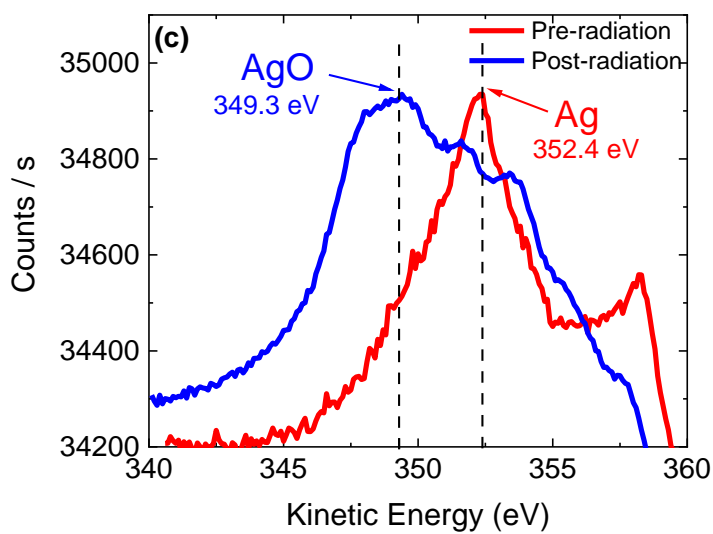

**Figure 3.** (a, b) XPS spectra for Si2p and O1s before (black line) and after (red line) radiation.

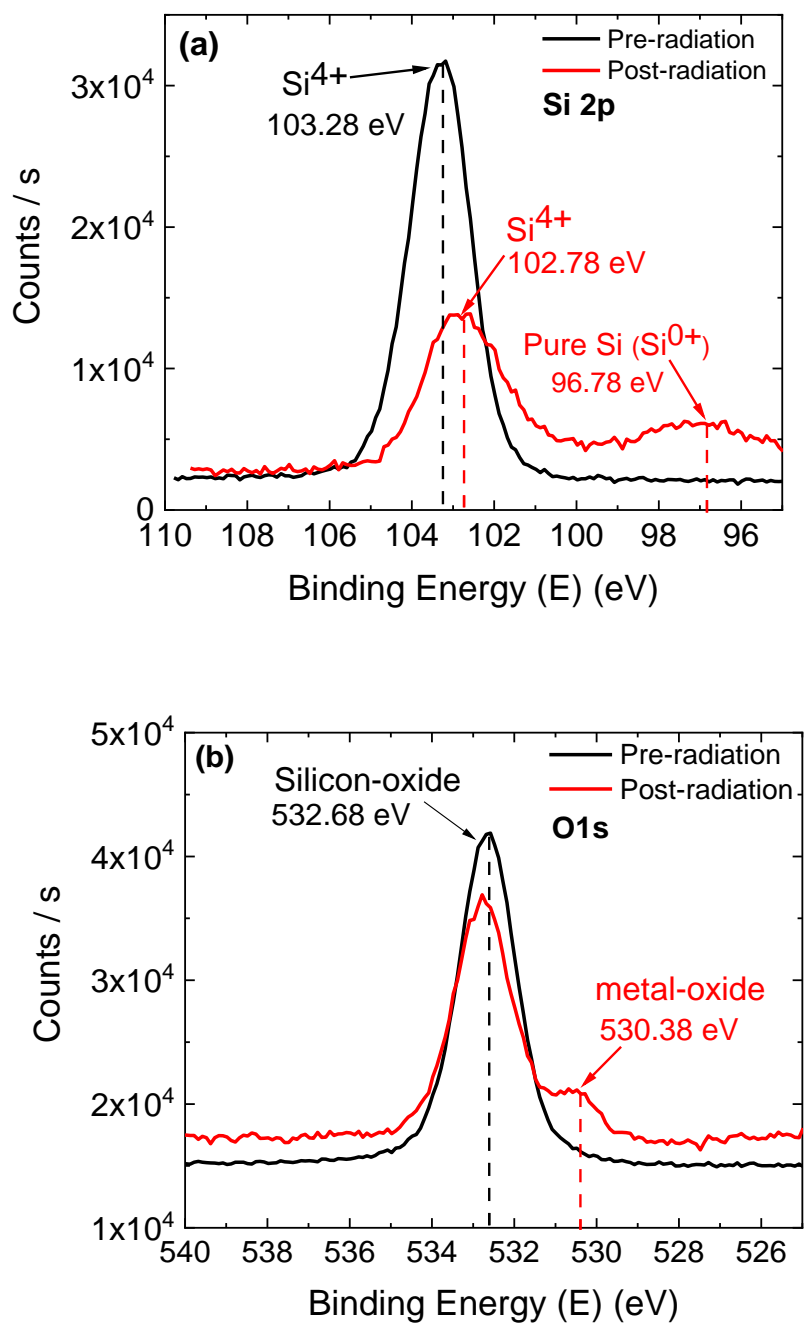

**Figure 4.** (a) Electrical circuit scheme for an artificial neuron with a diffusive memristor. (b) Experimentally observed voltage spikes measured across the memristor before (top, black) and after the irradiation (bottom, red).

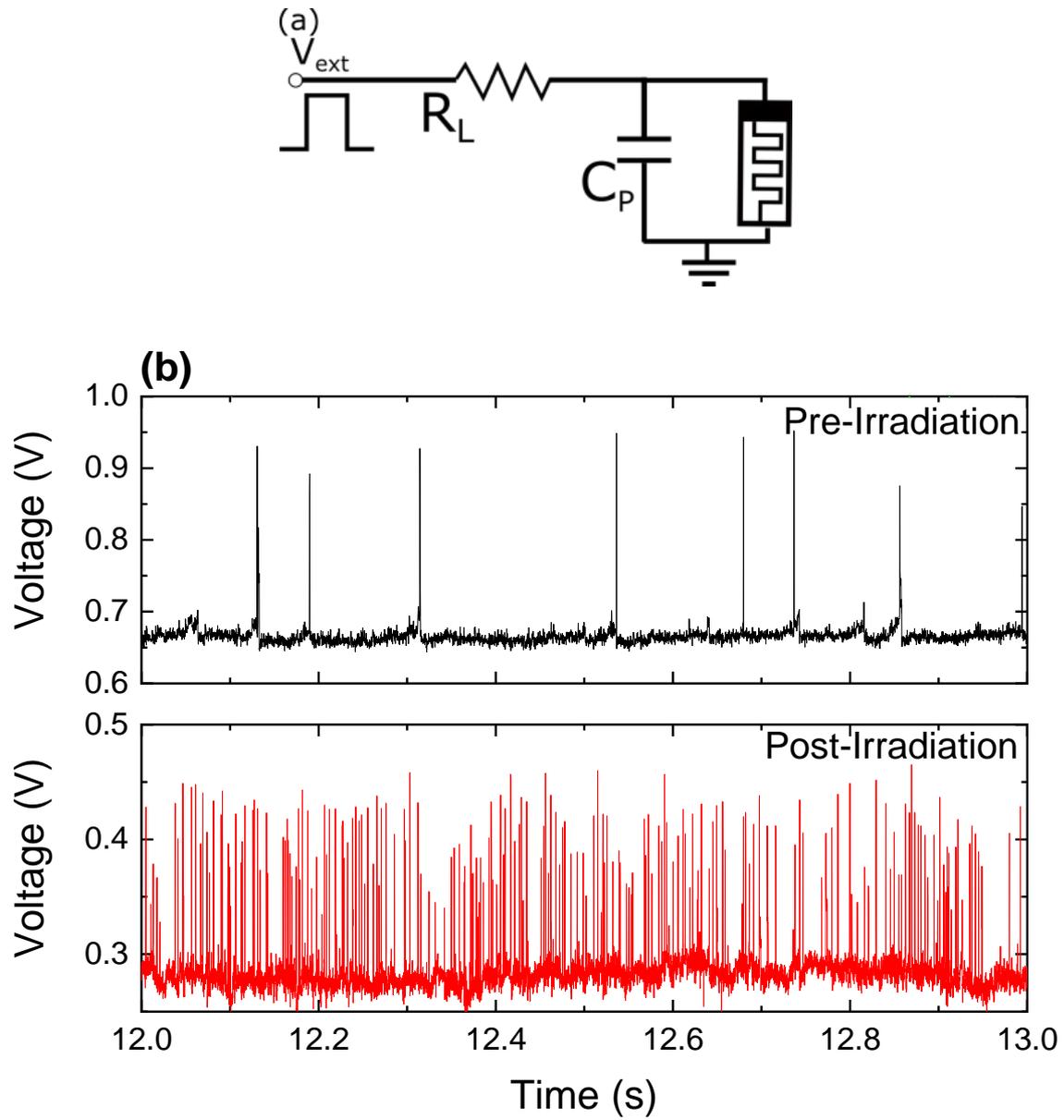

**Figure 5.** Schematics of a SiO$_x$:Ag diffusive memristor showing (a) Conduction filament in LRS for the pristine sample before irradiation. (b) Conduction filament in LRS for the sample after irradiation, containing Si NIs and oxygen vacancies.

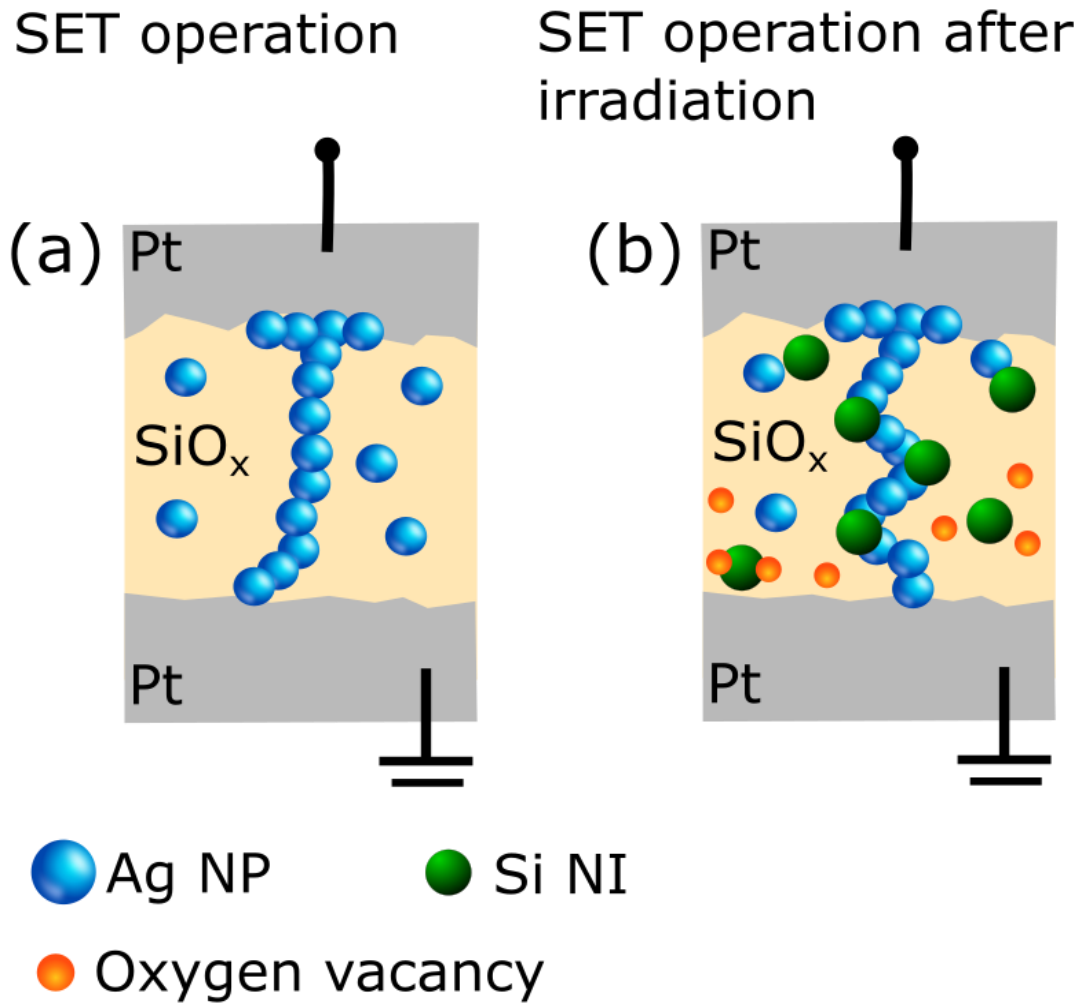

**Figure 6.** Simulated diffusion of four Ag clusters (black, red, green and blue curves) through a gap of size $L$ located between the electrodes of a diffusive memristor which is part of an artificial spiking neuron. The diffusion is of a 2D (a, c, e) or single-file (b, d, f) character. Spiking of conductance normalized to the maximal value (c, d) $x$ coordinate of a Ag cluster vs. time in the direction perpendicular to the electrodes (i.e. longitudinal diffusion) (e, f) $y$ coordinate of a Ag cluster vs time in the direction along the electrodes (i.e. transverse diffusion). The inset in (f) shows the potential $U(x)$ between the terminals.

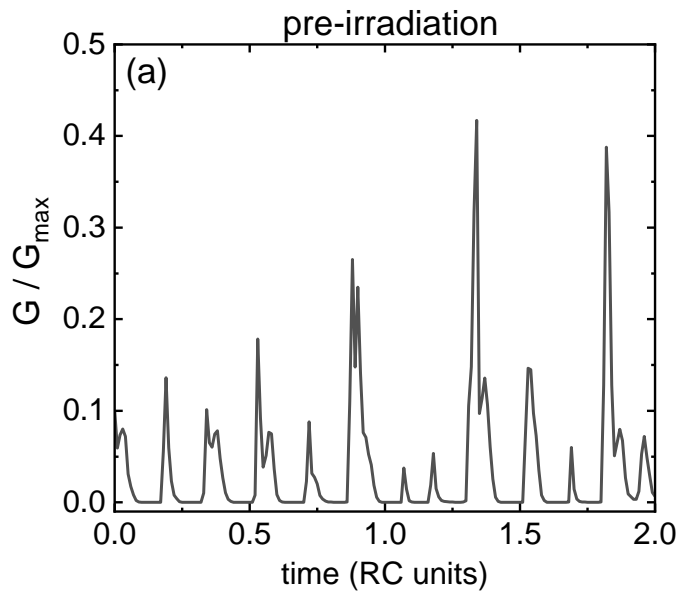
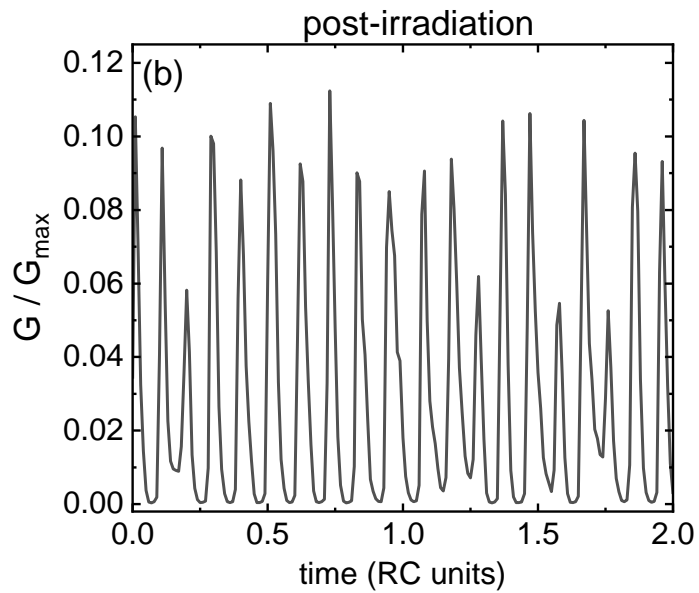
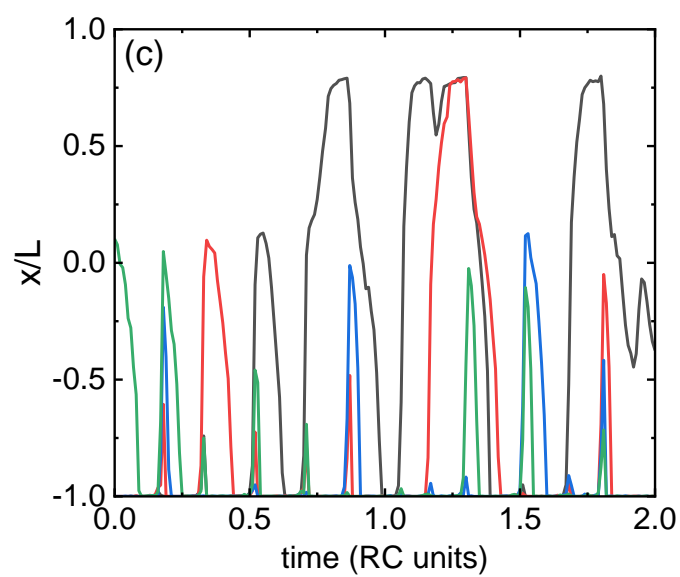
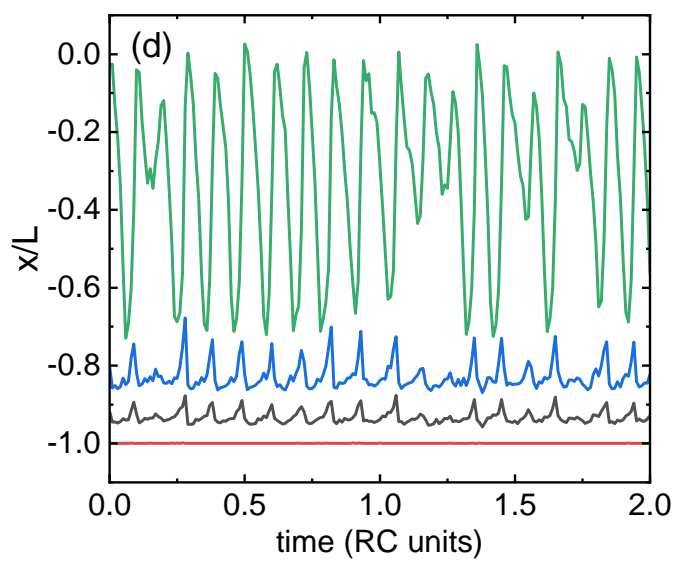
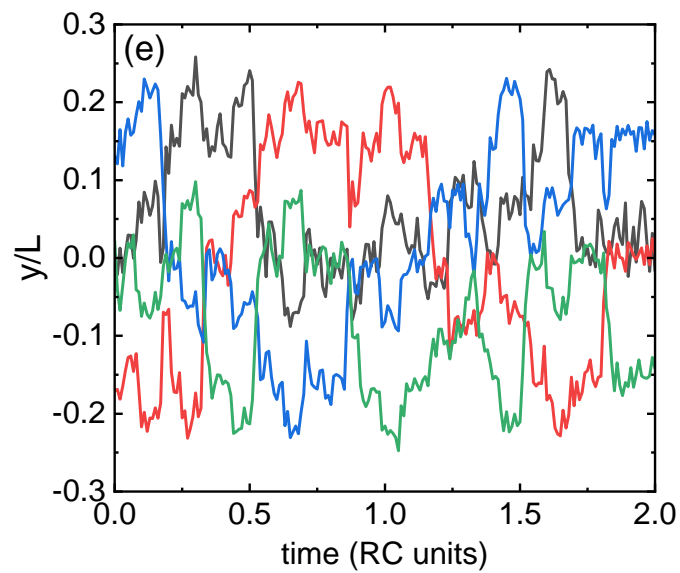
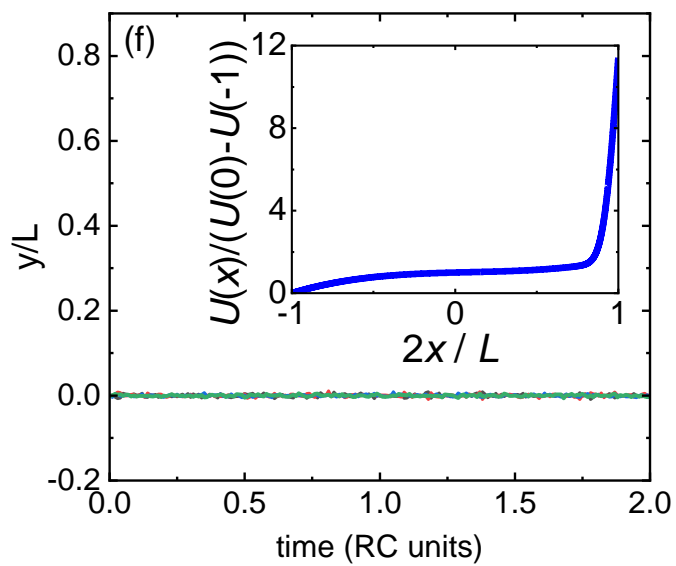